\documentclass{article} \usepackage{amsmath} \usepackage{amssymb}
\usepackage{epsfig}
\newcommand{\cC}{\mathcal{C}}

\newcommand{\cM}{\mathcal{M}}

\newcommand{\ie}{\textit{i.e. }}
\newcommand{\etc}{\textit{etc.}}

\font\openface=msbm10 at10pt

\def\Integers      {{\hbox{\openface Z}}}

\begin{document}
\title {Constructing an interval of Minkowski space from a causal set} \author{Joe
Henson} \maketitle
Institute for Theoretical
Physics, University of Utrecht, Minnaert Building, Leuvenlaan 4, 3584 CE
Utrecht, The Netherlands \footnote{Work also carried out at: Department of Mathematics, University of
California/San Diego, La Jolla, CA 92093-0112, USA. } .
\newpage
\begin{abstract}
A criticism sometimes made of the causal set quantum gravity program is that there is no practical scheme for identifying manifoldlike causal sets and finding embeddings of them into manifolds.  A computational method for constructing an approximate embedding of a small manifoldlike causal set into Minkowski space (or any spacetime that is approximately flat at short scales) is given, and tested in the 2 dimensional case.  This method can also be used to determine how manifoldlike a causal set is, and conversely to define scales of manifoldlikeness.
\end{abstract}
\vskip 1cm

In any quantum gravity program in which spacetime is replaced by a discrete structure, it is important to relate continuum approximations to the underlying structures, in order to relate known physics to the new theory (see e.g. \cite{Bombelli:2004si}).  In the causal set program \cite{SpacetimeAsCS}, the discrete/continuum correspondence relies on embeddings of the fundamental causal set (or \textit{causet} -- for definitions see \cite{SpacetimeAsCS,FirstSteps,Sorkin:2003bx}) into Lorentzian manifolds, and the concept of sprinkling.  A sprinkling is a random selection of points from the manifold made according to a Poisson process -- this means that the number of points sprinkled into any region depends only upon the volume $V$ of that region, according to
\begin{equation}
P(n)= \frac{V^n e^{-V}}{n!},
\end{equation}
where $P(n)$ is the probability of sprinkling $n$ elements into the region.  Natural units will be used throughout, and so this unit density Poisson process gives on average one element per Planck volume (which is here, for simplicity, assumed to be the fundamental discreteness scale).  A sprinkling defines a causal set, the elements being the sprinkled points, and the partial order being that induced on those points by the causal order of the Lorentzian manifold.  A manifold is considered to be an approximation to the causet if the causet could arise in this way with relatively high probability.  In this case the causet is said to be ``faithfully embeddable'' in the manifold.  In summary, number of elements corresponds to volume, while the partial order corresponds to continuum causal order.

Thus, the concept of faithful embedding not only gives a correspondence between causets and manifolds, but between elements of that causet and points in that manifold.  This enables us to speak of approximate distances between elements and so on.

This general idea of continuum approximation has been of great utility to the causal set quantum gravity program.  In particular, it is easy to investigate ``manifoldlike'' causets by carrying out sprinklings of a spacetime on computer (at least of conformally flat spacetimes), or relying on general properties of sprinklings to derive results.  This answers questions of the form ``given this (approximating) Lorentzian manifold, what are the properties of the underlying causet(s)?''.  See for example \cite{Dowker:2003hb,Dou:2003af,Ahmed:2002mj}.

However, the following criticism is often made of the idea: although it provides an in-principle correspondence between causets and manifolds, no practical scheme exists to find an embedding of a \textit{given} causal set into a manifold, or even to say whether it could be so embedded.  Indeed, the discrete/continuum correspondence principle outlined above cannot be used directly here.  Given a causal set, it would be highly impractical to compare it with various sprinklings until one was found which matched.  And since the causal set is posited as the fundamental structure (and moreover, since there are large numbers of non-manifoldlike causets), it is necessary to be able to identify those which are manifoldlike, and find embeddings for them, in a computationally tractable way.  For example, if a dynamics were to produce some typical output in computer simulations, it would be crucial to know if these could represent the manifold we observe.  While some necessary conditions are known (the matching of different dimension estimators at large scales \cite{FirstSteps,Reid:2002sj}, ``scale invariance'' \cite{Rideout:2000fh}, \etc) no necessary and sufficient condition for manifoldlikeness has been found.

Another problem with faithful embedding is that it may be too strict a condition.  By adding a small number of elements to any faithfully embeddable causet, it is possible to form a non-embeddable causet \cite{FirstSteps}.  If such ``closely similar'' causal sets are to have similar physical interpretations, the condition of faithful embedding would have to be relaxed (whether this is indeed the case depends on the final form of the dynamics, but in view of previous experience with quantum theories it seems very likely).  In this case it would be useful to have a scale of ``manifoldlikeness'' with faithfully embeddable causets as most manifoldlike.  Ideas of ``coarse-graining'' have previously been suggested to solve this problem \cite{FirstSteps}.

 
Progress towards these goals can be achieved with the application of the some remarkably simple ideas, in the form of a computer program.  When attempting to find an embedding for a causal set into a Lorentzian manifold, the causal set should somehow determine the curvature of that manifold.  But in the following computational scheme, the number of elements is practically limited, and 10000-element causal sets were tested.  The corresponding spacetime regions will therefore have a volume of approximately 10000 Planck volumes, and for regions of this size, curvature is negligible in all but the most extreme situations.  Therefore it is sensible, when dealing with relatively small causal sets, only to consider embeddings into Minkowski space at first, since deviations from this spacetime would generically be small.  The ideal would be to have a scheme that, when given a faithfully embeddable causal set $\cC$, took a region of Minkowski space $\cM$ (of the correct dimension) and assigned a point $x(e_i)$ to each element $e_i$ in $\cC$, such that $x(e_i) \subset J^-( x(e_j))$ in the causal order of $\cM$ iff $e_i \prec e_j$ in $\cC$, \textit{and} such that the distribution of assigned points in $\cM$ could have come from sprinkling with relatively high probability.  It would also be of use to have a scheme which only approximately does this.  If the points assigned to the elements of $\cC$ did not make a faithful embedding, but were approximately the correct distances from each other in the manifold, this would be just as useful for some purposes.

This letter describes a computational scheme to do this for small causal sets.  The scheme can be applied to any small ``causet interval'' (see below), and the level of success it has in producing a faithful embedding can be quantified and used as a measure of the ``manifoldlikeness'' of that causet.  Because the result only approximates a faithful embedding even when given a faithfully embeddable causet, the scheme can only approximately distinguish between faithfully embeddable causal sets and others; but if the yet-to-be-defined dynamics does not significantly distinguish either\footnote{this is perhaps an interesting question to address for the stochastic CSG dynamics \cite{CSG}.}, this is does not greatly affect the utility of the idea.  As yet, it has only been implemented to find 2D sprinklings, but would also be practical in other dimensions.

%

The scheme works as follows.  The $N$ elements of the causet $\cC$ will be called $e_0$,$e_1$,...,$e_{N-1}$.  $\cC$ is restricted to be an interval, \ie all elements lie causally between two elements, say $e_0$ and $e_1$.  As the name suggests, such causal sets correspond to causal intervals in the continuum.  Every element in $\cC$ is to be assigned co-ordinates in such a continuum interval.  In the 2D case light-cone co-ordinates $ds^2=2dudv$ will be used, and the co-ordinates assigned to $e_i$ will be called $(u_i,v_i)$.  The element $e_0$ is assigned the co-ordinates $(0,0)$.  While explaining the embedding procedure it is helpful to assume that $\cC$ is faithfully embeddable into 2D Minkowski space, and this will be so until otherwise noted.

The construction of an approximate embedding of $\cC$ has two stages.  Firstly, an assignment of position is made to each element using a well-known distance estimator for faithfully embeddable causal sets.  Due to the random nature of faithful embedding, these estimated position assignments are not perfect, but they are improved by a ``trial and error'' method at the second stage.

For the first stage, some way is needed to obtain manifold information from the causal set.  Fortunately, many simple methods of doing so already exist, and all that is required here is an estimator of distance between causally related elements, several of which are known \cite{FirstSteps,Brightwell:1990ha}.  The most useful for flat spacetimes (having a smaller variance than other methods in this case), involves  the number of elements causally between $e_x$ and $e_y$, $|I(e_x,e_y)|$, a number that approximates to the volume of the corresponding spacetime interval (it makes little difference if the end points are included in $I(e_x,e_y)$; they will be considered to be so here).  A simple calculation shows how the volume of the interval between two points is related to the distance between them in Minkowski space, giving
\begin{equation}
\label{2Ddist}
d(e_x,e_y) \approx \sqrt{2 |I(e_x,e_y)|},
\end{equation}
in the 2D case.

Because $|I(e_0,e_1)|=N$, $d(e_0,e_1) \approx \sqrt{2N}$, and so $e_1$ can be given the co-ordinates $(\sqrt{N},\sqrt{N})$.  Co-ordinates for the remaining elements can be almost completely determined by the distances between the element $e_i$ and the two elements $e_0$ and $e_1$.  A simple piece of geometry is employed to find $(u_i,v_i)$ :

\begin{align}
\label{geom}
d(e_1,e_i)^2 &=2(u_1-u_i)(v_1-v_i) \\
               &= d(e_0,e_1)^2 +d(e_0,e_i)^2 - 2(u_1v_i + u_iv_1),
\end{align}
(note that in these co-ordinates $d(e_0,e_i)=2u_iv_i$ ), giving

\begin{gather}
u_1v_i + u_iv_1 = X, \\
X= \frac{1}{2} \bigl( d(e_0,e_1)^2 + d(e_0,e_i)^2 - d(e_1,e_i)^2 \bigr) \\
\Longrightarrow \frac{1}{2} u_1 d(e_0,e_i) - X \, u_i + v_1 u_i^2 =0 \\
\label{e::u_i}
\Longrightarrow u_i=\frac{+X \pm \sqrt{X^2- d(e_0,e_1)^2 d(e_0,e_i)^2}}{2v_1},\\
v_i=\frac{d(e_0,e_i)^2}{2u_i}.
\end{gather}

In some cases, fluctuations in the distance measure could lead to complex co-ordinates.  In this case the complex part of the solution is thrown away, giving $u_i=v_i$.  The two possibilities left for the assignment of $(u_i,v_i)$ reflect the fact that not all the frame information has been fixed: the left-right symmetry still remains.  The first element placed after $e_0$ and $e_1$ fixes it, and after that all the other elements must be placed consistently with this choice.  In higher dimensions, a higher dimensional spherical symmetry would be fixed.

\begin{figure}[ht]
\centering
\resizebox{4.4in}{2.2in}{\includegraphics{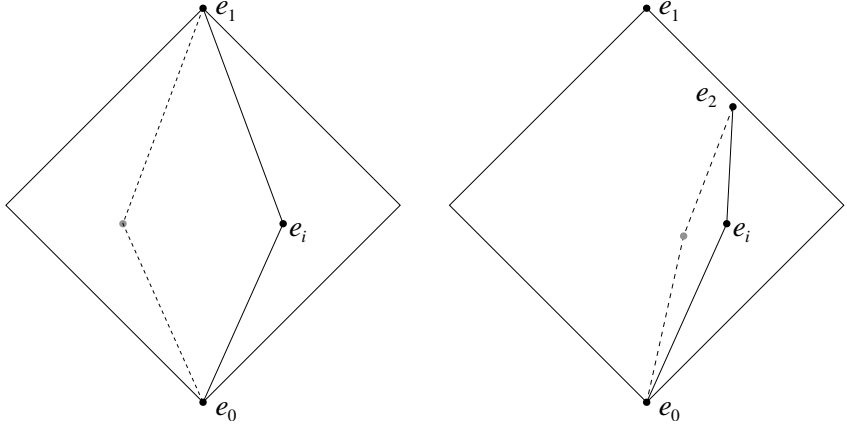}}
\caption{\small{
The process of assigning points in a Minkowski interval to elements in a causet interval.  The left hand diagram depicts an Interval of Minkowski.  The bottom and top elements $e_0$ and $e_1$ (represented by dots) can easily be assigned coordinates.  Timelike distances from these elements to another element $e_i$ can be estimated and a point assigned to it.  These distances fix the co-ordinates assigned to $e_i$ except for a $\Integers_2$ symmetry, as there are two points with the same distances from these points (as is shown with dashed lines and a grey dot).  The right hand diagram shows the distances from $e_i$ to $e_0$ and $e_2$.  This also gives two possible assignments of co-ordinates, but only one is consistent with the results shown on the left.
}\label{f:recon}}
\end{figure}

This last piece of co-ordinate freedom can be fixed by choosing a further element $e_2$, and arbitrarily placing it on one side of the interval.  The co-ordinates assigned to $e_2$ are now known, and all other assigned points must now be placed consistently with this choice.  Take an element $e_i$ which is causally related to $e_2$.  It has been shown how, given the co-ordinates assigned to two elements, and the estimated distances from them, a pair of possible positions for a third element (causally between the two) can be found - in particular $e_0$ and $e_1$ were used to place $e_i$.  In the same way $e_2$ and $e_0$ (or $e_1$) can be used to find a second possible pair of sites for $e_i$.  One site of each pair should approximately match, revealing the co-ordinates at which to place $e_i$.  This is illustrated in figure \ref{f:recon}.

Not all elements will be causally related to $e_2$.  However, those that \textit{are} related to $e_2$ can be assigned co-ordinates.  From these an element $e_3$ is chosen, which is used in the same way as $e_2$ to place all elements causally related to it, and so on.  In practice only a few well-chosen ``reference elements'' like $e_2$ and $e_3$ are needed to determine the co-ordinates of all elements.  Rare exceptions are elements sprinkled so close to the corner of the interval that they are only related to $e_1$ and $e_0$, in which case there is no way to tell if the element should be at near $(0,\sqrt{N})$ or $(\sqrt{N},0)$.

In this 2D case, two pairs of points must be used to completely fix the co-ordinates assigned to an element, and in general $n$ pairs would be needed in $n$ dimensions.  This is because the interval in $n$ dimensions has a $n-1$ spherical symmetry which must be fixed from the reference points.  For example, in 3D, using $d(e_i,e_0)$ and $d(e_i,e_0)$ only restricts possible values of $x(e_i)$ to a circle; these distances tell us nothing about the angle about the axis of the interval at which to place $x(e_i)$, and so a circular symmetry remains to be fixed.  Now consider the distance to $e_i$ from some reference element $e_2$.  Two points on the circle have the correct distance from $e_2$.  A further reference point would be needed to determine which is the correct position to assign to $e_i$.

A rough assignment of co-ordinates to each element using the timelike distance estimator has now been made.  To gauge the effectiveness of the scheme in approximating a faithful embedding of $\cC$, some measures of success are now introduced.  As well as the original causet $\cC$, we now also have the partial order on these points induced by the causal order of the Minkowski interval, which will be called $\cC'$.  These two are not necessarily the same, even if $\cC$ is faithfully embeddable; the scheme is quite rough and does not use all the information available in $\cC$ to place the elements.  Let the proportion of relations in $\cC$ incorrectly matched in $\cC'$ be called $R$.  In simulations, $R \approx 1.7\%$.  This is an average taken over 1000 runs, in which 10000-element faithfully embeddable causets were used as input to the scheme (as are the other such results reported below).

This number $R$ only contains causal information: it tells us how close $\cC'$ is to $\cC$, but nothing more about whether $\cC$ can be \textit{faithfully} embedded into $\cM$.  To answer this, the embedding has to be tested in some way to find whether it could have come from a sprinkling with relatively high probability.  Ideally, $\chi$-squared tests would be used to determine this as in \cite{Reid:2002sj}.  In the program, as a rough measure, the standard deviation of the number of elements in each of 100 sub-regions of the interval was measured.  The absolute difference between this and the expected value for sprinklings is here called $S$.  In simulations $S \approx 6.1$.  Setting an acceptable maximum for $S$ amounts to a practical definition of ``relatively high probability of sprinkling'' in the description of faithful embedding.  Other measures of success can be imagined, but $R$ and $S$ were found to be easy and fast to compute in this first, ``prototype'' implementation.

These statistics are used in the second stage of the scheme, at which the assignments of co-ordinates are improved.  The assigned points are each moved by small distances in order to minimise $R$.  In practice the best new position was selected from a random list, and the process repeated a number of times until improvements ceased.  Although the value of $R$ is not sensitive to volume information, this idea is easy to implement, and was found to significantly improve results.  On average, after this refinement $R \approx 0.34\%$ and $S \approx 0.70$.  In no runs of the program could $R$ be reduced to exactly 0, furnishing a true embedding of the causet into the Minkowski interval;  hopefully, further developments on the basic ideas given here could achieve this.

These measures of closeness to faithful embedding have another use.  When applied to a causal set $\cC$ that is not faithfully embeddable, the scheme cannot achieve such low values of $R$ or $S$, because there is no assignment of co-ordinates to the elements that perfectly matches the causal relations in $\cC$ and could have come from sprinkling.  But causal sets which achieve scores close to the minimum possible scores could be considered as close to faithfully embeddable.  Thus, on the one hand the scheme can be used, with a limited degree of certainty, to identify faithfully embeddable causets.  Sprinklings of 2D causet intervals could be tested with the scheme to find the distribution of values for $R$ and $S$ for faithful embeddings, and these values compared to those from any given causet.  If a given causal set $\cC$ achieved values of $R$ and $S$ that could have come from a faithfully embeddable causal set with relatively high probability, the scheme would accept $\cC$ as faithfully embeddable.  Any improvements to the embedding scheme (producing lower values of $R$ and $S$ where possible) would also improve the accuracy with which faithfully embeddable causets can be identified.  On the other hand, the tolerances for values of $R$ and $S$ could be set lower, allowing a more relaxed critrion of manifoldlikeness: one that only required the causal set to be ``close'' to faithfully embeddable (this closeness being measured by the valuses of $R$ and $S$ achieved).  In other words, the standard idea of faithful embeddability gives the strictest bounds on acceptable ``manifoldlike'' values of $R$ and $S$, but it is possible to imagine generalising the criterion of manifoldlikeness by relaxing those bounds.

The question of what values of $R$ and $S$ should be considered as acceptable (\ie indicative of manfoldlikeness) is difficult and will not be dealt with here, for two main reasons.  Firstly, a relaxation of the criterion of faithful embedding will probably be required by the dynamics, but on the other hand we do not want to go too far: one causal set should not be approximated by two very different manifolds.  How much relaxation will be required by the dynamics, and how much can be introduced without spoiling the discrete-continuum correspondence, is not decided as yet. Secondly the values of $R$ and $S$ obtained when $\cC$ is a sprinkling of a 2D interval depend on the details of the scheme; since the simple method used here can no doubt be improved upon (hopefully reducing $R$ to 0 with high probability), there is little point in laying down a standard at this stage.  Similarly, better estimators of random sprinkling than $S$ might be preferred if more computational power was available.  Having said that, this scheme does demonstrate the possibility of distinguishing non-manifoldlike causal sets.  For example, a regular lattice produces $S \approx 10$, and tentative results show that causal sets produced by sprinkling with variable density give values for $R$ and $S$ that are clearly distinguishable from values given by causets that are faithfully embeddable into 2D Minkowski.  More detailed results on such topics are left for future work.

What has been achieved by the invention of the scheme?  Previously, there existed the definition of faithful embedding, and this gave an in-principle criterion by which to associate manifolds with casual sets.  It is important to note that faithful embeddability (into a given manifold) is the property of the causal set itself, not of a particular embedding.  So for each causal set, this criterion gives a yes/no answer to the question ``is this manifold a good approximation to this causal set?''.  The problem is that this in-principle property of the causal set is difficult to determine; the criterion is defined in terms of sprinkling candidate manifolds.  But we do know practical ways to answer questions of the form ``Assuming that this causet \textit{does} faithfully embed into 2D Minkowski, what is the approximate distance between these two (timelike related) elements?''  The scheme presented here uses a distance estimator to apply the standard discrete/continuum correspondence ``in reverse''.  It gives the first practical way to distinguish (with limited certainty) between manifoldlike and non-manifoldlike causets, and construct an approximate embedding for the causal set, using only the distance estimator and elementary geometry.  It also suggests a way to relax the condition of faithful embeddability by using ``estimators of success'' like $R$ and $S$ to define scales of manifoldlikeness.

The calculations involved are computationally practical (and curvature issues are avoided) only with small numbers of elements; however, presented with a larger causal set, several small subcausets could be taken and checked for manifoldlikeness.  There are many possible extensions of the idea.  Implementation in larger dimensions would be desirable, requiring an extension of the simple geometrical results given above.  With this done, the dimension of the causal set $\cC$ could first be estimated with one of the many available methods \cite{FirstSteps}, before passing to the appropriate embedding routine.

Although very simple in reasoning and implementation, to the author's knowledge this method has not been used before.  This is probably because it is not obvious that the estimation of timelike distance would be accurate enough to give a reasonable assignment of co-ordinates to each element.  This has now been shown to be the case for small causets that are faithfully embeddable into 2D Minkowski.  The purpose of the scheme was to attempt to construct a faithful embedding without use of any given embedding; but, if the causet $\cC$ was indeed created by sprinkling (as it was in simulations), the scheme's assigned co-ordinate values can be compared to those from the original sprinkling.  The average discrepancy in the $u$ co-ordinate is only $\approx 0.82$ fundamental units before the refinement and $\approx 0.37$ units after (an average over all 1000 runs, and over all elements).  No scheme could possibly reduce this discrepancy to $0$ (because many slightly different sprinklings can produce the same causal set), so this is a good result -- the co-ordinate ``uncertainty'' is below the fundamental length in the natural frame of the interval.  This gives an example of how much manifold information can be present in even a small causal set.

It would be interesting to know how the scale of manifoldlikeness associated with this scheme compares to previous suggestions of the ``course-graining'' type, and whether this can be seen as a special case of those ideas.  It would also be very desirable to know of some simple properties of the causal set that are indicative of manifoldlikeness.  This would avoid the need for reliance on computational power inherent in the scheme presented here, and may give a more natural way to test large causal sets with curved approximating manifolds.  Hopefully as more conditions for manifoldlikeness become known, they will combine into one that is necessary \textit{and} sufficient.

The author is grateful to Rafael Sorkin and Sumati Surya for correspondence and
discussions of this work.  The author was supported by DARPA grant
F49620-02-C-0010R at UCSD, where some work on the article was carried out.

\bibliographystyle{h-physrev3}
\bibliography{first}
\end{document}